# Study of neutron multiplicity in $^{232}$Th (n,f) reaction using TALYS - 1.96

Punit Dubey and Ajay Kumar
*Banaras Hindu University, Varanasi, India*
E-mail: ajaytyagi@bhu.ac.in

**Introduction:**

The nuclear scientific community views $^{232}$Th as an option for fuel in the future nuclear energy program. The fissile element utilized in contemporary commercial reactors is uranium-235 ($^{235}$U), which constitutes a mere 0.72% of natural uranium. The $^{238}$U isotope, which constitutes 99.27% of natural uranium, cannot undergo fission within existing thermal reactors. To optimize the utilization of available resources, two primary fuel cycles are suggested for prospective implementation: the $^{238}$U/$^{239}$Pu and $^{232}$Th/$^{233}$U fuel cycles. In the context of the $^{238}$U/$^{239}$Pu cycle, the fissile material is identified as $^{239}$Pu, which is derived from $^{238}$U by the process of neutron capture followed by two successive beta decays. In the $^{232}$Th/$^{233}$U cycle, the fissile material is $^{233}$U, which undergoes transmutation from $^{232}$Th through the processes of neutron capture and two subsequent beta decays. The $^{232}$Th/$^{233}$U cycle presents appealing characteristics, including the greater natural abundance of thorium compared to uranium, the generation of a lower quantity of waste, and a reduced presence of transuranic isotopes in the waste. The generation of the fissile nucleus $^{233}$U in the thorium-uranium fuel cycle occurs through the process of the $^{232}$Th(n,γ)$^{233}$Th reaction, which is then followed by two consecutive β-decays. The cross-section of the $^{232}$Th(n, 2n) $^{231}$Th reaction exhibits a sharp increase beyond a threshold energy of 6.648 MeV. The provided diagram illustrates the schematic representation of the Thorium-Uranium (Th–U) fuel cycle.

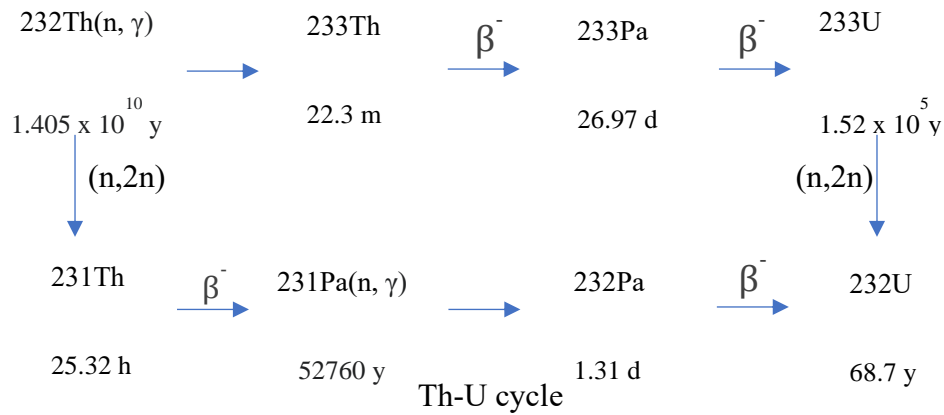

Th-U cycle

Several experimental studies have been undertaken to ascertain the cross-section [1-4]; yet there is a scarcity of investigations [5-7] that have been carried out to compute the overall neutron multiplicity above an energy threshold of 10 MeV. Several scholars have conducted theoretical and experimental investigations on detecting neutron multiplicities in fusion-fission dynamics generated by heavy ions [8–14]. However, works [5–7] are scarce, focusing on reactions induced by particles and neutrons.

In this study, we have performed calculations to determine the neutron multiplicity for the $^{232}$Th(n,f) reaction at various incident energies. These calculations were conducted using the TALYS 1.96 [15].

**Result and Discussion:**

In this work, we have compared the experimental data of average neutron multiplicity at different incident energies from EXFOR with the evaluated data from ENDF/B-VI, JENDL-4.0, and the calculated data from TALYS-1.96, as shown in Fig. 1. The experimental data are in good agreement with the evaluated data from both the ENDF/B-VI and JENDL-4.0 libraries and at high incident energy (14.7 MeV), the TALYS 1.96 data are also in agreement with the experimental data.

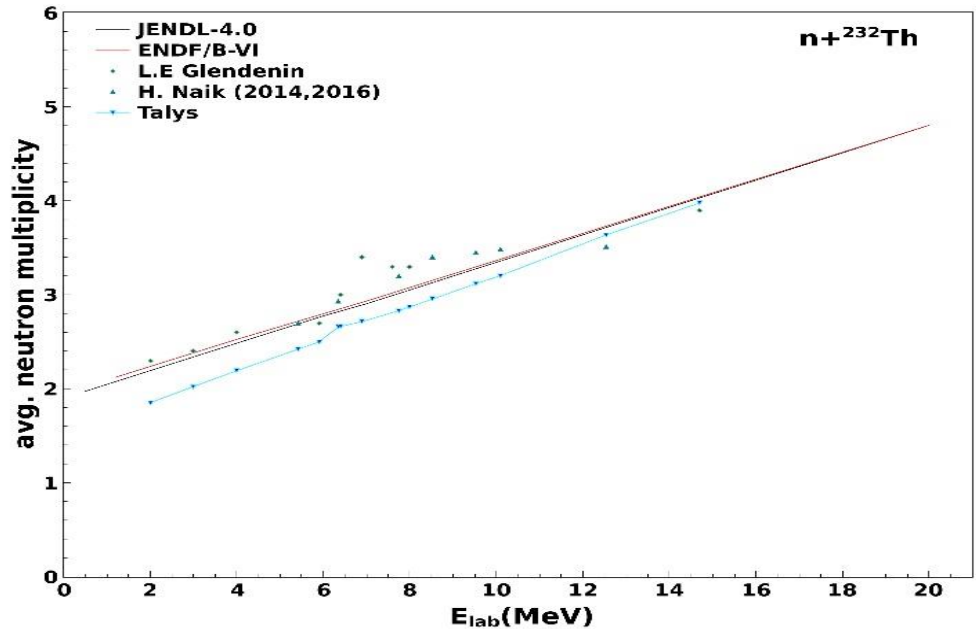

Fig.1: Comparison of experimental and evaluated data w.r.t TALYS calculated data for $^{232}$Th (n,f) reaction.

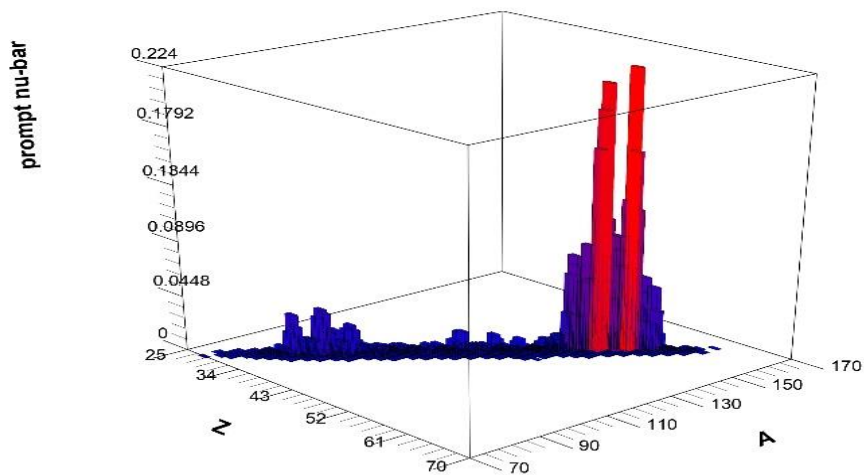

Fig.2: Prompt neutron multiplicity as a function of mass (A) and charge (Z) using TALYS 1.96 for $^{232}$Th (n,f) reaction at 14.7 MeV.

As shown in Fig. 2, we have also calculated the prompt neutron multiplicity at 14.7 MeV as a function of mass (A) and charge (Z). This figure reveals that neutron multiplicity is highly dependent on both Z and A fission fragments, as evidenced by the peaks at $_{54}Xe^{141,142,143,144}$ and $_{56}Ba^{146,147,148,149}$. Therefore, it is evident from the graph that neutron multiplicity varies with respect to Z and A.

The theoretical framework being suggested facilitates the computation of some measurable neutron characteristics, specifically the neutron emission function ν(A), which represents the total number of released neutrons as a function of the initial fragment mass and the total neutron emission number ñ. The aforementioned functions of neutrons play a crucial role in the field of neutron physics and find extensive applications in various contexts involving neutron fluxes. The most significant factors influencing ν(A) and ñ include the isotopic makeup of the initial nucleus and its excitation energy or temperature T. The calculation of the ν (A) function is derived from the probabilistic determination of the realisation (yield) of a two-fragment cluster, which consists of a pre-neutron emission fragment with mass A and an equilibrium number of neutrons ñ.

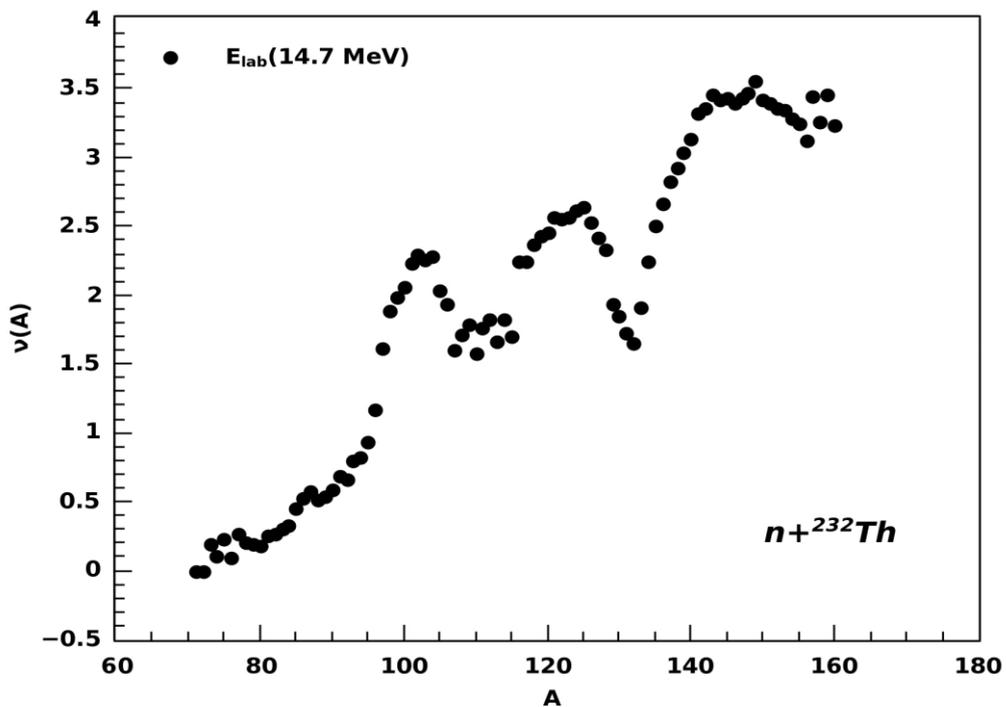

Fig.3: The fission neutron yield is given as a function of the fission fragment mass (A) $^{232}$Th (n,f) calculated using TALYS-1.96 code at 14.7 MeV.

We have also calculated the neutron emission functions for fission fragments of $^{233}$Th using TALYS 1.96, as can be seen in Fig. 3. As one can see, the theoretical calculations show a "saw-tooth"-curve of the neutron multiplicity, namely the peak at about 102, minimum in the vicinity of 110, further growth in the range of 126 and decrease to 133 and another peak at 150 and then a decrease.

## Conclusion:

At a high excitation energy (14.7 MeV), the current study shows that the computed neutron multiplicity values from TALYS and the experimental data are very close to each other. To comprehensively understand the impact of neutron multiplicity, it is essential to acquire additional data from experiments at higher values of incident energies. Consequently, we have plans to conduct a forthcoming investigation to address this research gap.

**Acknowledgement:** The author (Punit Dubey) is grateful to the Prime Minister Research Fellowship (PMRF ID - 0101841) for the financial support for this work. One of the authors (A. Kumar) would like to thank the Institutions of Eminence (IoE) BHU [Grant No. 6031] and IUAC-UGC, Government of India (Sanction No. IUAC/XIII.7/UFR-71353).

## References:

[1] Yonghao Chen, Yiwei Yang, Zhizhou Ren, Wei Jiang, Ruirui Fan, Han Yi, Rong Liu *et al.,* Physical Letters B **839**, 137832 (2023).
[2] Yu M. Gledenov, Zengqi Cui, Jie Liu, Haoyu Jiang, Yiwei Hu, Haofan Bai, Jinxiang Chen *et al.,* Eur. Phys. J. A **58**, 86 (2022).
[3] Rita Crasta, H. Naik, S. V. Suryanarayana, B. S. Shivashankar, V. K. Mulik, P. M. Prajapati, Ganesh Sanjeev *et al.,* Annals of Nuclear Energy **47**, 160-165 (2012).
[4] Sadhana Mukerji, H. Naik, S. V. Suryanarayana, S. Chachara, B. S. Shivashankar, V. Mulik, Rita Crasta *et al.,* Pramana-Journal of Physics **79**, 249-262 (2012).
[5] L. E. Glendenin, J. E. Gindler, I. Ahmad, D. J. Henderson, and J. W. Meadows Phys. Rev. C **22**, 152 (1980).
[6] H. Naik, Rita Crasta, S.V. Suryanarayana, P.M. Prajapati, V.K. Mulik, B.S. Shivasankar, K.C. Jagadeesan, S.V. Thakare, S.C. Sharma, and A. Goswami, The European Physical Journal A, **50**, 144 (2014).
[7] H. Naik, Sadhana Mukherji, S.V. Suryanarayana, K.C. Jagadeesa, S.V. Thakare, S.C. Sharma, Nuclear Physics A, **952**, 100-120 (2016).
[8] N. K. Rai, A. Gandhi, Ajay Kumar, N. Saneesh, M. Kumar, G. Kaur, A. Parihari, D. Arora, K. S. Golda, A. Jhingan, P. Sugathan, T. K. Ghosh, Jhilam Sadhukhan, B. K. Nayak, Nabendu K. Deb, S. Biswas, and A. Chakraborty, Phys. Rev. C **100**, (1), 014614 (2019).
[9] N. K. Rai, A Gandhi, M. T. Senthil Kannan, S. K. Roy, N. Saneesh, M. Kumar, G. Kaur, D. Arora, K. S. Golda, A. Jhingan, P. Sugathan, T. K. Ghosh, Jhilam Sadhukhan, B. K. Nayak, Nabendu K. Deb, Saumyajit Biswas, A. Chakraborty, A. Parihari, and Ajay Kumar, Journal of Physics G: Nuclear and Particle Physics **49**, (3), 035103 (2022).
[10] N. K. Rai, Vivek Mishra, and Ajay Kumar, Phys. Rev. C **98**, 024626 (2018).
[11] A. Kumar, A. Kumar, G. Singh, B.K Yogi, R. Kumar, S.K. Datta, M.B. Chatterjee, and I.M. Govil, Phys. Rev. C **68**, 034603 (2003).
[12] Ajay Kumar, A. Kumar, G. Singh, Hardev Singh, R.P. Singh, Rakesh Kumar, K.S Golda, S.K. Datta, and I.M.Govil, Phys. Rev. C 70, 044607 (2004).
[13] A. Kumar, H.Singh, R. Kumar, I.M. Govil, R.P.Singh, Rakesh Kumar, B.K.Yogi, K.S Golda, S.K Datta, and G. Viesti, Nucl. Phys.A **798**, 1 (2008).
[14] Ajay Kumar, A. Kumar, B.R. Behra, Hardev Singh, R.P. Singh, R. Kumar and K.S. Golda, EPJ Web of Conferences **86**, 00019 (2015).
[15] Arjan Koning, D. Rochman, Nucl. Data Sheets **113**, 2841 (2012).